\documentclass[conference,compsoc]{IEEEtran}

\usepackage{graphicx}
\usepackage[labelformat=simple]{subcaption}
\usepackage{changepage}
\usepackage{multirow}
\usepackage{rotating}
\usepackage{pifont}
\usepackage{tcolorbox}
\usepackage{array}
\usepackage{fontawesome5}
\usepackage{makecell}
\usepackage{hhline} 
\usepackage{svg}
\usepackage{float}
\usepackage{balance}
\usepackage{booktabs}
\usepackage{tcolorbox}
\usepackage{enumitem}
\usepackage{url}
\usepackage{xurl}
\usepackage[hidelinks]{hyperref}

\newcommand{\parheading}[1]{\textbf{{#1}}}
\newcommand{\eg}{\textit{e.g.,}}
\newcommand{\ie}{\textit{i.e.,}}

\newcommand{\etal}{et al.}
\newcommand{\wrt}{w.r.t.}

\tcbset{colback=gray!5, colframe=black!80, boxrule=0.5pt, arc=1.5mm, left=2mm, right=2mm, top=1mm, bottom=1mm}

\newcommand{\kidsmode}{{Kids Mode}}
\newcommand{\tiktok}{{TikTok}}
\newcommand{\tkm}{{TKM}}
\newcommand{\tiktokapi}{{TikTok-Api}}
\newcommand{\fyp}{{FYP}}

\newcommand{\totobsall}{{1471}}
\newcommand{\totmatchall}{{1438}}
\newcommand{\totobsneutral}{{535}}
\newcommand{\totmatchneutral}{{516}}
\newcommand{\totobsexp}{{936}}
\newcommand{\totmatchexp}{{922}}
\newcommand{\totaggunique}{{485}}

\newcommand{\dataset}{{\tkm{} Labeled Content Dataset}}
\newcommand{\experiments}{{Account Age and Username Experiments}}
\newcommand{\neutralcoll}{{Neutral Data Collection}}

\ifCLASSOPTIONcompsoc
  \usepackage[nocompress]{cite}
\else
  \usepackage{cite}
\fi

\begin{document}
\title{When Kids Mode Isn't For Kids: Investigating TikTok's ``Under 13 Experience''}

\author{\IEEEauthorblockN{Olivia Figueira,
Pranathi Chamarthi,
Tu Le, 
Athina Markopoulou
}
\IEEEauthorblockA{University of California, Irvine}
}

\maketitle

\begin{abstract}
TikTok, the social media platform that is popular among children and adolescents, offers a more restrictive ``Under 13 Experience'' exclusively for young users in the US, also known as TikTok's ``Kids Mode''. While prior research has studied various aspects of TikTok's regular mode, including privacy and personalization, TikTok's Kids Mode remains understudied, and there is a lack of transparency regarding its content curation and its safety and privacy protections for children. In this paper, (i) we propose an auditing methodology to comprehensively investigate TikTok's Kids Mode and (ii) we apply it to characterize the platform's content curation and determine the prevalence of child-directed content, based on regulations in the Children's Online Privacy Protection Act (COPPA). We find that 83\% of videos observed on the ``For You'' page in Kids Mode are actually {\em not} child-directed, and even inappropriate content was found. The platform also lacks critical features, namely parental controls and accessibility settings. Our findings have important design and regulatory implications, as children may be incentivized to use TikTok's regular mode instead of Kids Mode, where they are known to be exposed to further safety and privacy risks.
\end{abstract}

\IEEEpeerreviewmaketitle

\section{Introduction}\label{sec:intro}

\tiktok{} is a popular social media platform known for its short-form video content and has amassed over 1.6 billion monthly users worldwide~\cite{iqbal_business_tiktok_revenue_2025}. \tiktok{} is particularly popular among children and adolescents in the United States (US)~\cite{anderson_pew_teens_socialmedia_2023, zhong_nyt_child_tiktok_2020}.
As the platform's dominance has grown, \tiktok{} has also been scrutinized for its potential harms to young users, such as addictive behaviors stemming from the highly personalized stream of content on the ``For You'' page (\fyp{}), impacts to self-esteem from video filters, dangerous viral trends and challenges, exposure to inappropriate content, and privacy risks due to the collection and sharing of personal information~\cite{nyag_tiktok_lawsuit_2024, herzlich_nyp_tiktok_meditation_2025, allyn_npr_addiction_known_2024, ftc_tiktok_lawsuit_2024}, leading to lawsuits and regulatory enforcement actions~\cite{nyag_tiktok_lawsuit_2024, ftc_tiktok_lawsuit_2024}.

\textbf{\tiktok{}'s Underexplored \kidsmode{}.} Prior research has investigated \tiktok{}'s content recommendations and personalization factors~\cite{le_autolike_2025, kaplan_tiktokaudit_2024, boeker_tiktok_factors_2022, hilbert_bigtech_2024}, inappropriate content~\cite{balat_tikguard_2024, milton_mental_health_content_2023}, users' perceptions toward \tiktok{} content~\cite{milton_mental_health_content_2023, schluchter_mental_health_tiktok_2024}, and data privacy and compliance concerns for young users~\cite{figueira_diffaudit_2024, hilbert_bigtech_2024}. However, another version of \tiktok{} aimed towards young users remains understudied: the ``Under 13 Experience''~\cite{tiktok_kidsmode_2025}. Available exclusively in the US, \tiktok{} provides a more restrictive ``Under 13 Experience'' for young users, which we refer to as \tiktok{}'s Kids Mode (\tkm{})~\cite{tiktok_kidsmode_2025, ftc_tiktok_lawsuit_2024}. \tkm{}\footnote{Once the \tiktok{} app is opened, the user is prompted to enter their date of birth. If the entered date results in an age younger than 13, the app automatically diverts the user to the \tkm{}-specific account creation process, which only requires the user to enter a username and a password.} is an extremely limited version of \tiktok{}, presumably in an effort to shield young users from potentially risky behaviors and interactions. \tkm{} users can only ``like'' or report videos on the \fyp{} (\ie{} no viewing or posting comments, no sharing, and no video descriptions), users cannot post any videos since their account is private, users cannot search for any content, and users cannot message any other users or view any other users' profiles. Thus, it would seem that \tkm{} aims to protect young users from the potential harms of \tiktok{}'s regular mode, and \tiktok{} advertises that they have partnered with a third-party service to curate content specifically for \tkm{}~\cite{tiktok_kidsmode_2025}, however the details remain unclear. 

\textbf{Research Problem.} \tiktok{}'s documentation regarding \tkm{} lacks details about how they curate \tkm{} content. For \tkm{} to be a safe and privacy-protecting platform for children under 13, it should be free from inappropriate content and should contain safety and privacy features. 
Due to a settlement with the Federal Trade Commission (FTC) over alleged Children’s Online Privacy Protection Act (COPPA) violations~\cite{ftc_youtube_content_rules_2019}, YouTube videos must indicate whether the content is child-directed following COPPA's definition of child-directed content and services. 
\tkm{} content should also be specifically directed to children in alignment with COPPA's definition of child-directed content~\cite{ftc_youtube_content_rules_2019, ftc_coppa_faq_2020}. Otherwise, children may not be interested or engaged and, as a result, may be incentivized to abandon \tkm{} for \tiktok{}'s regular mode. In the latter case, children would have to lie about their age (\ie{} \tiktok{}'s regular mode is restricted to 13 years and older) and may be exposed to well-documented risks such as mental health harms, inappropriate content, targeted advertising, and data privacy risks~\cite{figueira_diffaudit_2024, balat_tikguard_2024, milton_mental_health_content_2023, hilbert_bigtech_2024}. 
To that end, we aim to comprehensively investigate \tkm{} and answer the following research questions (RQs):
\begin{itemize}
    \item{\textbf{RQ1:} How can \tkm{}'s content curation be characterized (\ie{} what categories of content are shown to children in \tkm{}, how prevalent are \tkm{} video authors, is there frequent video repetition)?} 
    \item{\textbf{RQ2:} Does \tkm{} contain child-directed content, inappropriate content, or advertisements?}  
    \item{\textbf{RQ3:} What are the safety and privacy design implications of \tkm{} as a service?}
\end{itemize}

\textbf{Contributions.} To address these RQs, we make two contributions: (1) we design and implement a methodology to programmatically collect and analyze \tkm{} content, which has not been done before, and (2) we apply our methodology to \tkm{} and report our findings, as further described next.

First, \textbf{we develop a \tkm{} auditing methodology} for collecting and analyzing \tiktok{} videos shown on the \fyp{} of \tkm{} on an Android mobile device. \tkm{} is extremely limited \wrt{} its features compared to \tiktok{}'s regular mode, and thus prior works' \tiktok{} crawling approaches are unsuitable for \tkm{}. To overcome these challenges, we develop a novel approach to crawling \tkm{} videos with fewer app-specific dependencies, enabling programmatic \tkm{} data collection and experiments. Our methodology enables auditors and researchers to independently investigate \tkm{} content without direct cooperation from \tiktok{}.
We also develop custom content labels to analyze whether \tkm{} videos are child-directed, based on the COPPA definition~\cite{coppa_rule_2013}, and whether \tkm{} videos contain inappropriate content and advertisements. We apply these custom content labels by manually labeling all video observations in our study, as we describe next.

Second, \textbf{we apply our \tkm{} auditing methodology} to collect and analyze \tkm{} content, resulting in a \dataset{} with \totobsall{} total \tkm{} video observations, all of which we manually labeled with our custom content labels, and our dataset contains \totaggunique{} unique videos (\eg{} there are repeated observations) from 443 video authors. We make the following observations: First, the majority of \tkm{} content is actually {\em not child-directed,} which is at odds with the purpose of \tkm{} as a service. Based on our manual labeling of child-directed content, according to COPPA's definition of child-directed content~\cite{coppa_rule_2013}, we found that 83\% of the unique \tkm{} videos we observed were not child-directed.
Second, we also observed inappropriate content in \tkm{}, including sexually explicit and profane content. While there were not many instances (\ie{} 9/485), \tkm{} should not contain \textit{any} inappropriate content because it is intended for children. 
Third, we found that \tkm{} is repetitive: we observed frequent video repetition across our dataset, and 24\% of the videos were repeated at least twice, potentially indicating a limited content inventory for \tkm{}, which is surprising given that \tiktok{}'s regular mode is rife with child-directed content.
Finally, by analyzing \tkm{}'s features, we found that it lacks critical safety and privacy features, including meaningful parental controls and accessibility settings.

Based on these findings, we present recommendations for children and parents, as well as for children's service providers and regulators. We implore \tkm{} and other service providers, who are considering developing child-specific social media platforms, to scrutinize their systems and content curation processes. If a service is advertised to be specifically for children, then it should be child-directed, age-appropriate, and contain effective parental controls and usability features. Otherwise, the purpose of such a service is questionable: \tkm{} currently appears to be ``for kids'' only by name. \tkm{} does not actually provide content that appeals to kids, thus incentivizing them to switch to \tiktok{}'s regular mode, where they can find ample child-directed content but also be exposed to well-known risks. 
Regulators and lawmakers can utilize our findings and methodology to investigate child-directed services based on their content and privacy practices. 
Our methodology and findings can also be used to investigate other research questions\footnote{We plan to publicly release our dataset and software artifacts.}, beyond the scope of this paper, such as to enable user studies regarding social media content appropriateness and child-directedness, as well as for the development of automated child-directed content classifiers. Further, our methodology can be applied to study how \tkm{} content changes over time as well as personalization factors, such as the impact of different interactions with \tkm{} videos.

\parheading{Outline.} 
The rest of this paper is structured as follows: Section~\ref{sec:background_related_work} presents background information and related work. Section~\ref{sec:methodology} discusses our \tkm{} auditing methodology, including our data collection and labeling process (Section~\ref{subsec:method_kidsmode_framework}) and experimental design (Section~\ref{subsec:method_experiments}). Section~\ref{sec:findings} presents our findings. Section~\ref{sec:discussion} discusses the implications of our findings and recommendations for service providers and regulators. Finally, Section~\ref{sec:conclusion} concludes the paper.

\section{Background and Our Work in Perspective}\label{sec:background_related_work}

\subsection{The Children's Online Privacy Protection Act}\label{sec:background}

This section presents how the Children's Online Privacy Protection Act (COPPA)~\cite{coppa_rule_2013} motivates the design of our \tkm{} auditing methodology and discusses relevant enforcement actions against children-directed services.

\subsubsection{COPPA and Child-Directed Services}\label{subsec:backg_coppa}

COPPA provides privacy protections for children under 13 years of age in the US and applies to websites and online services, or parts thereof, that are specifically \textit{directed to children}, and COPPA defines what constitutes a child-directed service based on many factors (\eg{} ``use of animated characters or child-oriented activities and incentives'', ``presence of child celebrities or celebrities who appeal to children''\footnote{16 C.F.R.§312.2 ``Web site or online service directed to children'' (1)}). This definition also applies to content on a website or online service, such as social media videos~\cite{ftc_youtube_content_rules_2019}. If an online service or its content are deemed to be child-directed, then they must comply with COPPA, which includes obtaining verifiable parental consent for the collection and sharing of data about children under 13 beyond that which is necessary for functionality of their service. 
If an online platform claims to be directed to children, such as \tkm{}, then we expect its content and visuals to be clearly intended for children, considering the factors in COPPA's definition.

\subsubsection{Enforcement Actions}\label{subsec:backg_enforcement}

The Federal Trade Commission (FTC) regularly conducts investigations to enforce COPPA regulations. Most relevant to this paper are the FTC's recent lawsuits against YouTube and \tiktok{} for alleged COPPA violations. 
In a 2019 settlement with YouTube, the FTC alleged that while YouTube claimed their platform was not child-directed, they had internal knowledge that there were children on their platform watching videos that were directed to children while also targeting advertisements to these viewers without parental consent~\cite{ftc_youtube_lawsuit_2019}.
As a result, the FTC mandated that YouTube introduce a system requiring channel owners to mark content as child-directed, using COPPA's definition of child-directed content~\cite{coppa_rule_2013, ftc_youtube_content_rules_2019}, and YouTube was prohibited from placing targeted advertisements on such videos~\cite{ftc_youtube_lawsuit_2019}. 
Similarly, in a 2024 lawsuit, the FTC alleged that \tiktok{} knew there were millions of children on \tiktok{}'s regular mode (\ie{} restricted to users 13 years and older) but did not remove their accounts and continued to collect and share children's personal information without parental consent~\cite{ftc_tiktok_lawsuit_2024}. The FTC also claimed that \tkm{} illegally collected and shared children's personal information without parental consent and made it unnecessarily complicated  for parents to submit data privacy requests on behalf of their children. 

While we do not specifically study privacy violations in the context of COPPA in this work, we aim to expand on these previous investigations of \tiktok{} by characterizing \tkm{}'s content curation and analyzing the prevalence of child-directed content, using COPPA's definition of child-directed content. If \tkm{} content is not child-directed, then the purpose of such a service is questionable and should be further audited by enforcement entities.

\subsection{Related Work}\label{sec:related_work}
In this section, we present the related literature on children's online privacy and safety as well as content auditing and moderation on social media. We then discuss the scope and contribution of our work in comparison.

\subsubsection{Children's Online Privacy and Safety}\label{subsec:related_child_priv_safety}

Children are vulnerable to various types of online privacy and safety risks, ranging from invasive data collection to cyberbullying, online harassment, and harmful content~\cite{jun_zhao_i_2019, kaiwen_sun_they_2021, ybarra_youth_2004, twyman_comparing_2010, chou_whisper_2024}. Prior work has shown that many apps designed for children, in various domains such as mobile~\cite{irwin_reyes_wont_2018}, smart toys~\cite{gordon_chu_security_2019}, and voice assistants~\cite{tu_le_skillbot_2022}, violated privacy regulations by improperly collecting and sharing personal data. Other work has also revealed that parental control apps could introduce excessive surveillance and insecure data handling~\cite{suzan_ali_betrayed_2020, alvaro_feal_angel_2020}. Additionally, advertising directed toward children and adolescents could lead to inappropriate content and tracking~\cite{moti_tracking_2024, tinhinane_medjkoune_marketing_2023, marisa_meyer_advertising_2019, xiaomei_cai_advertisements_2008, xiaomei_cai_online_2013}. Recently, the advancement of generative artificial intelligence has also introduced various risks to children~\cite{yu_exploring_2025}.

Existing literature on children’s interactions with the digital world is often grounded in user studies that examine how children and their parents perceive privacy and navigate privacy-related decisions~\cite{cami_goray_youths_2022, cao_understanding_2024, jun_zhao_i_2019, kaiwen_sun_they_2021, leah_zhang-kennedy_nosy_2016, priya_kumar_no_2017}. Notably, children struggle to understand certain types of privacy risks~\cite{jun_zhao_i_2019, priya_kumar_no_2017}. They also often rely on their parents for privacy settings~\cite{priya_kumar_no_2017}. However, children are concerned about being monitored by parents as they described feelings of being stalked and negative perceptions of parental control features~\cite{digitalwellnesslab_safety_2023, arup_kumar_ghosh_safety_2018}. Previous research also studied usage of parental control mechanisms and parental oversight strategies for adolescents and children~\cite{peiyi_yang_towards_2023, mamtaj_akter_parental_2022, maria_grazia_lo_cricchio_parental_2022, pamela_wisniewski_parental_2017, phoebe_k_chua_what_2021, kuzminykh_howmuch_2019, williams_youth_2023}. While parents are concerned about children's privacy, they often compromise their children's privacy or help them lie about their age online to gain access to age-restricted services, suggesting that parents may not always be reliable to protect children online~\cite{hargittai2011parents, minkus2015children}.

\subsubsection{Content Auditing and Moderation}\label{subsec:related_content_mod}

Prior studies have developed frameworks for content auditing and moderation to identify safety and privacy issues, such as~\cite{jiang_tradeoff_2023, jhaver_personalizing_2023, kumar_watch_2024, kolla_llm_2024, scheuerman_severity_2021, schaffner_community_2024, vaccaro_contestability_2021, gomez_algo_2024}. However, very few focused specifically on investigating content made for children and adolescents. In particular, recent studies have investigated the effectiveness of content moderation systems for children's videos~\cite{ahmed_potential_2023}, explored designs of content moderation tools that consider children's needs~\cite{sas_informing_2023}, and examined how content classification processes affect children's safety on social media platforms such as YouTube~\cite{ma_labeling_2024}. Other studies have looked into identifying risky content delivered to children in voice assistants~\cite{tu_le_skillbot_2022}, targeted advertising practices directed toward children~\cite{moti_tracking_2024, khan_adexposure_2024, tinhinane_medjkoune_marketing_2023}, and how personalization systems shape children's content exposure in harmful ways~\cite{hilbert_bigtech_2024}. Researchers have also designed privacy guidelines and protections for young users~\cite{siew_yong_risk_2011, john_dempsey_children_2022, max_van_kleek_better_2017, liccardi_can_2014, ge_wang_12_2023, ge_wang_informing_2022, priya_kumar_co_designing_2018} and investigated the challenges developers face when creating child-directed apps~\cite{anirudh_ekambaranathan_money_2021, ekambaranathan_how_2023}.

\subsection{Scope and Contribution}\label{subsec:related_scope}
Specifically for \tiktok{}, several studies have proposed methodologies for auditing content recommendations and personalization on \tiktok{}'s regular mode~\cite{kaplan_tiktokaudit_2024, le_autolike_2025, boeker_tiktok_factors_2022} and law compliance of both \tiktok{}'s regular mode and \tkm{} \wrt{} data collection and sharing~\cite{figueira_diffaudit_2024}. Prior work found that \tiktok{}'s regular mode shows more child-directed content to users that behave like a child based on their app interactions~\cite{hilbert_bigtech_2024}, demonstrating that it is feasible to infer a user's age based on their behavior.
Regarding child-directed content analysis, the closest work to ours is \cite{hilbert_bigtech_2024}, however they only used a single label in their manual labeling process to determine whether a video is child-directed or not. In contrast, we develop a more detailed labeling approach with four labels based on COPPA's definition of child-directed content~\cite{coppa_rule_2013, ftc_youtube_content_rules_2019}. Additionally, \cite{figueira_diffaudit_2024} studied \tkm{} in particular, but they focused on COPPA and CCPA compliance \wrt{} data collection and sharing, not \tkm{} content.

To the best of our knowledge, there is no prior comprehensive auditing approach that facilitates child-directed content analysis for child-directed services. \tkm{}, in particular, remains understudied. We bridge this gap by developing a novel content auditing methodology designed for child-directed services and we apply it to \tkm{}, enabling us to investigate content curation on \tkm{} and the prevalence of child-directed content, as guided by COPPA~\cite{coppa_rule_2013, ftc_youtube_content_rules_2019}.

\begin{figure*}[ht!]
    \centering
    \includegraphics[width=0.95\textwidth]{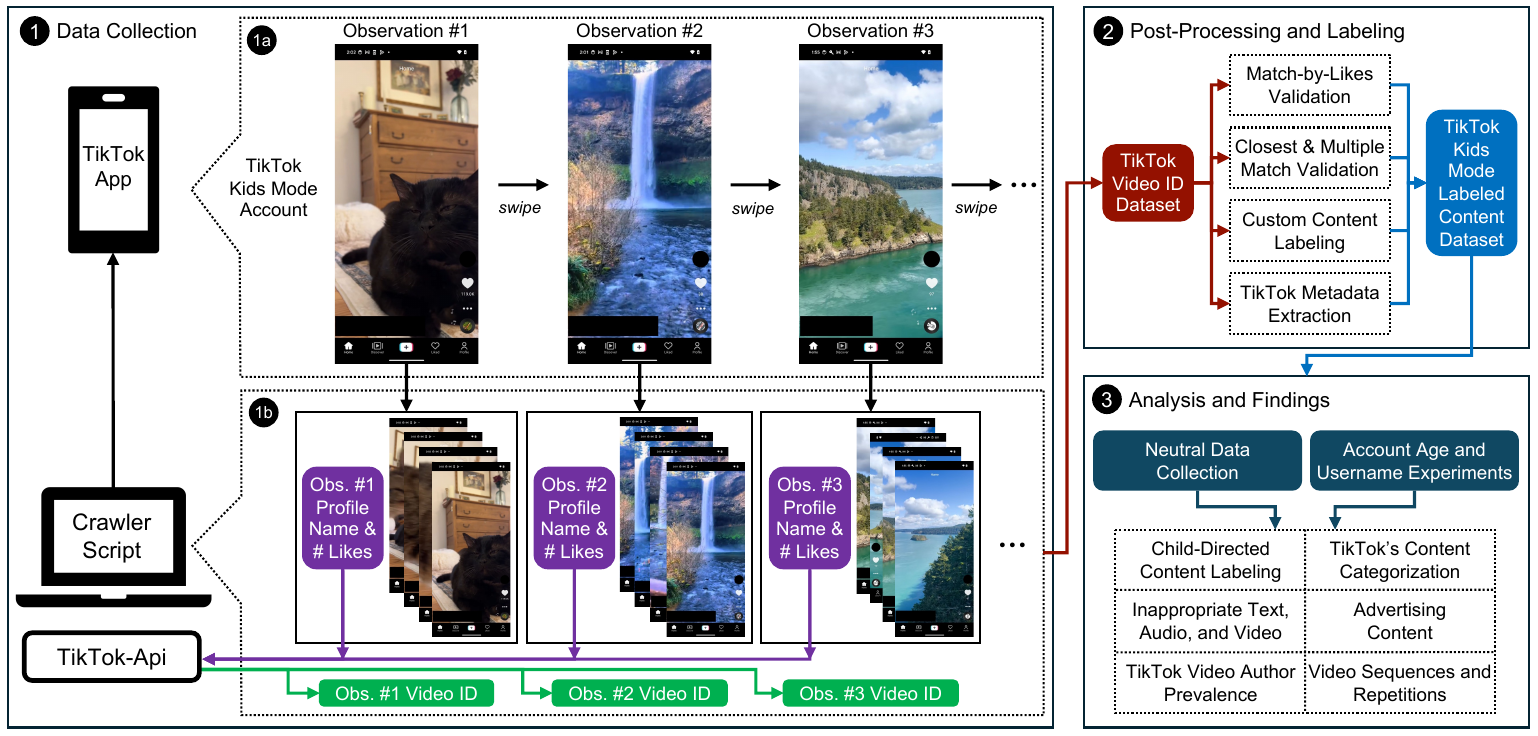}
    \begin{minipage}{\linewidth}
      \caption{
      \textbf{\tiktok{}'s \kidsmode{} Auditing Methodology Overview.} \small{
      \textbf{(1) Data Collection}: Our crawler script initiates and controls the crawl of \tiktok{} videos on the mobile device with the \tiktok{} app installed. \textbf{(1a)} For each video observed, the crawler will \textbf{(1b)} record the unique profile name and number of likes and capture screenshots of representative scenes in the video. The crawler uses the \tiktokapi{}~\cite{teather_tiktok-api_2025} to find the video on the video author's profile using the number of likes at the time of observation (``Obs.''), and otherwise will record all exact matches or the closest match, which comprise the \textit{\tiktok{} Video ID Dataset}. \textbf{(2) Post-Processing and Labeling} involves \textbf{\textit{Match-by-Likes Validation}} and \textbf{\textit{Closest and Multiple Match Validation}} to ensure match validity, \textbf{\textit{Custom Content Labeling}} using our set of child-directed, advertising, and inappropriate content labels, and \textbf{\textit{\tiktok{} Metadata Extraction}} through HTTP requests with the validated video URLs. \textbf{(3) Analysis and Findings}: with our final \textit{\dataset{}}, we analyze various axes, including custom content labeling, content categories, video author details, and account age/username experiments.}
      }
      \label{fig:auditing_methodology}
    \end{minipage}
\end{figure*}

\begin{figure}[ht!]
    \centering
    \includegraphics[width=0.49\textwidth]{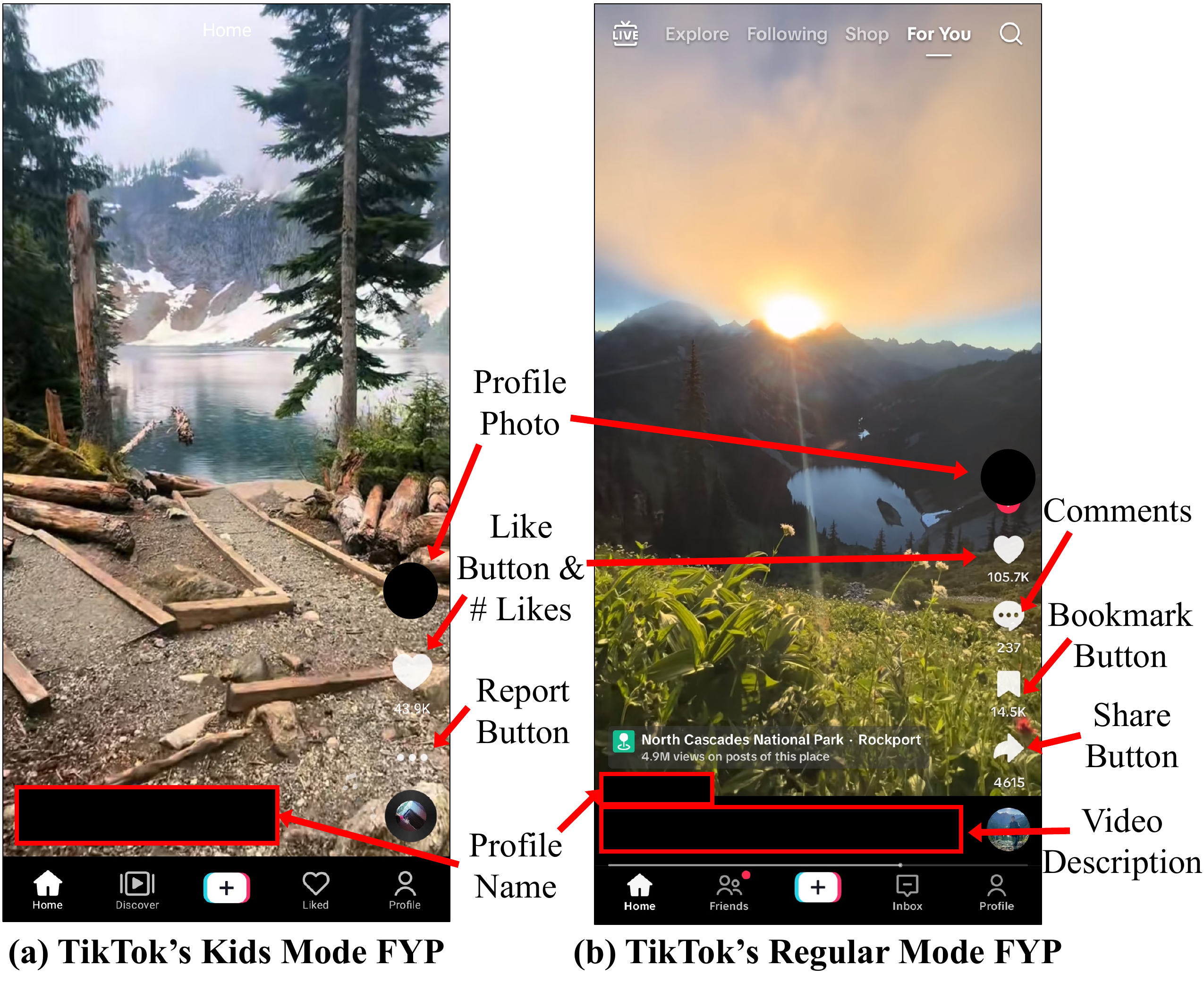}
    \begin{minipage}{\linewidth}
      \caption{
      \textbf{\tiktok{}'s \kidsmode{} and \tiktok{}'s Regular Mode ``For You'' Page (\fyp{}) Interfaces.} \small{Figure (a) presents a screenshot from the \tkm{} \fyp{} and indicates the features and information provided (\ie{} ``Like'' button, number of likes, ``Report'' button, and the profile name and photo of the video's author.) Figure (b) presents a screenshot from \tiktok{}'s regular mode and indicates the additional features provided (\ie{} comments, ``Bookmark'' button, ``Share'' button, and the video's description). Identifying information has been redacted.}
      }
      \label{fig:tiktok_ui_screenshots}
    \end{minipage}
\end{figure}

\section{Methodology}\label{sec:methodology}

In this section, we discuss the details of our auditing methodology (Section~\ref{subsec:method_kidsmode_framework}) and the experimental design guiding our data collection (Section~\ref{subsec:method_experiments}). An overview is shown in Figure~\ref{fig:auditing_methodology}.

\subsection{\tiktok{}'s \kidsmode{} Auditing Methodology}\label{subsec:method_kidsmode_framework}

\subsubsection{Data Collection}\label{subsubsec:method_data_collection}

Our data collection pipeline includes a Python program on a computer connected via the Android Debug Bridge~\cite{adb} to an Android mobile device with the \tiktok{} app installed. The program automatically creates user accounts and collects videos shown on the \fyp{} of \tkm{}, enabling us to conduct experiments and construct datasets of \tkm{} content.

\parheading{Approach on \tiktok{}'s Regular Mode.} Prior work has studied \tiktok{}'s regular mode to investigate content recommendations and personalization~\cite{le_autolike_2025, kaplan_tiktokaudit_2024, boeker_tiktok_factors_2022}. They employed the UIAutomator2 library~\cite{uiautomator2} to simulate user interactions, such as swiping and clicking on elements. UIAutomator2 can be utilized to reveal the XML representation of the devices' screen and identify user interface elements, such as the ``Like'' and ``Share'' buttons on each \tiktok{} video. Clicking the ``Share'' button reveals a URL of the video, which enables the collection of video observations. By making an HTTP request with this URL, we can extract more metadata about the video, including the full, unique numerical identifier (hereinafter referred to as ``video ID'') for that video, which is typically 19 characters in length. This video ID is critical to distinguishing between videos and extracting their corresponding metadata, such as the video author's information, \tiktok{}'s content categories, and video engagement statistics (\eg{} likes, shares, comments).

\parheading{Challenges.} A key difference between \tiktok{}'s regular mode and \tkm{} is that there is no ``Share'' button in \tkm{}, as shown in Figure~\ref{fig:tiktok_ui_screenshots}, and thus the video URL cannot be directly obtained from the app's user interface. This presents a challenge to our data collection goal, which cannot be solved using existing methods from prior work.
Importantly, the only \textit{information} available on a \tiktok{} video in \tkm{} is the video author's profile name (which is a unique ID for each user), the video author's profile photo, the number of likes, and the title of the audio playing in the video. The only \textit{features} available are the ``Like'' button and a button to report the video, such as for inappropriate content (\eg{} see Figure~\ref{fig:tiktok_ui_screenshots}). \tiktok{}'s regular mode also allows users to view and post comments, share the video, and more. In contrast, \tkm{} is extremely limited in both functionality and information. 
Additionally, users have the same limitations when using \tkm{} in a browser, and the URLs of \tiktok{} videos shown in the browser do not correspond to their actual, unique video IDs in \tiktok{}'s regular mode. This seems to be due to \tkm{}'s content selection and delivery system, as the videos selected for \tkm{} are served from specific content delivery networks and their URLs differ from the actual URLs on \tiktok{}'s regular mode.

\parheading{Our Solution.} We first attempted to use network traffic decryption techniques (\eg{} PCAPdroid) to identify the video URLs while using \tkm{} on the mobile device. However, the resulting traffic neither contained the video URLs nor the unique video IDs corresponding with the videos observed in the app.
Consequently, we develop an alternative approach\footnote{Our solution can also be easily applied to \tiktok{}'s regular mode.} (depicted in Figure~\ref{fig:auditing_methodology}, Step 1b) to obtain the video ID for each video that is shown in \tkm{}: (1) we extract the author's unique profile name and number of likes from the user interface for a given video with UIAutomator2~\cite{uiautomator2}, and then (2) we employ the open-source \tiktokapi{}~\cite{teather_tiktok-api_2025} to find the same video on the author's profile page by matching the number of likes in real time.

Through testing our solution, we found that the number of likes observed for a video in \tkm{} may vary from the actual number of likes on the same video, and sometimes the original video no longer exists on \tiktok{} (\eg{} it may have been deleted) but it still shows up on \tkm{}. These discrepancies seem to arise from \tiktok{}'s content selection and delivery system, which may cause \tkm{} videos to be out of date compared to the actual videos in \tiktok{}'s regular mode. Thus, if we are unable to find any videos that match based on the number of likes observed in \tkm{}, we try to find the video that has the closest number of likes, of which there may be ties, and we refer to these cases as ``closest-match-by-likes''. There may also be cases in which there are multiple videos on the author's profile with the same number of likes as the video observed in \tkm{} (\ie{} our target number of likes), and thus we collect all such cases, and we refer to these as ``multiple-match-by-likes''.

We then develop a method to capture screenshots of representative scenes in each video to later match against the potential video match(es) identified during the data collection. For each video observation in \tkm{}, we capture screenshots for 30 seconds and compute frame similarity scores via OpenCV~\cite{opencv} to detect scene changes in the video. With this approach, we obtain a set of screenshots that serve as representative scenes from the video, and we use them in our post-processing and labeling steps (detailed in Section~\ref{subsubsec:method_postprocess} and depicted in Figure~\ref{fig:auditing_methodology}, Step 2) to validate whether the screenshots of the video observed in \tkm{} match the video identified during the data collection.

\subsubsection{Post-Processing and Labeling}\label{subsubsec:method_postprocess}
Our data collection solution requires post-processing to validate that the video collected from the author's profile matches the one observed in \tkm{} and to conduct content labeling to prepare our dataset for analysis (Figure~\ref{fig:auditing_methodology}, Step 2).

\parheading{Validation and Labeling Method.}
Two researchers independently conduct (1) match-by-likes validation (\ie{} compare the representative scene screenshots with the \tiktok{} video identified during the crawl), (2) closest-match-by-likes validation (\ie{} confirm whether the closest matching video(s) by likes identified during the crawl matches the representative scene screenshots), (3) multiple-match-by-likes validation (\ie{} identify which of the videos matched with our target number of likes matches the representative scene screenshots), and (4) custom content labeling (\ie{} we label the final identified videos with our custom content labels), which we discuss next. After each step, the researchers discuss and finalize the labels for the videos until full consensus is reached.

\begin{table*}[t!]
    \centering
    \small
    \caption{\textbf{Data Collection Experimental Design and Dataset Summary. }\small{This table lists the experimental design and dataset summary for our \neutralcoll{} (``Neutral'' column in the table) and our \experiments{} (``E\#C\#'', denoting the experiment (E) and category (C) per experiment), which together comprise our \dataset{}. We report dataset statistics, including the number of \tkm{} videos observed in each experiment and how many of those observed were successfully matched (\ie{} video ID was identified). For the Neutral Data Collection, we use a single neutral account and the same device, denoted as ``AN''. For the \experiments{}, each experiment utilizes three accounts with varying age and username (with a gendered name). See Section~\ref{subsec:method_experiments} for more details. The rightmost columns includes the totals across the columns. Note that the total number of unique videos (485) is not the sum of the final row, rather the number of unique videos across our \dataset{}, which includes both our Neutral and \experiments{} datasets.}}
    \resizebox{\textwidth}{!}{
    \begin{tabular}{p{3.75cm} c || ccc | ccc | ccc || ccc | ccc | ccc || c ||}
        & \multicolumn{1}{c}{\textbf{Neutral}} & \multicolumn{3}{c}{\textbf{E1C1}}  &  \multicolumn{3}{c}{\textbf{E1C2}} &  \multicolumn{3}{c}{\textbf{E1C3}} &  \multicolumn{3}{c}{\textbf{E2C1}} & \multicolumn{3}{c}{\textbf{E2C2}} & \multicolumn{3}{c}{\textbf{E2C3}} & \multicolumn{1}{c}{\textbf{Total}}\\
         \midrule
         
        \textbf{Account/Device}   &  \multicolumn{1}{||c||}{AN} & A1 & A2 & A3 & A1 & A2 & A3 & A1 & A2 & A3 & A1 & A2 & A3 & A1 & A2 & A3 & A1 & A2 & A3 & 19 \\
        \textbf{Account Username}     &  \multicolumn{1}{||c||}{User} & Noah & Noah & Noah & Emma & Emma & Emma & User & User & User & Olivia & Liam & User & Emma & Liam & User & Sophia & Noah & User & 19 \\
        \textbf{Account Age}      &  \multicolumn{1}{||c||}{7} & 3 & 7 & 11 & 3 & 7 & 11 & 3 & 7 & 11 & 3 & 3 & 3 & 7 & 7 & 7 & 11 & 11 & 11 & 3 \\
        \midrule

        \textbf{\# Videos Observed}      &  \multicolumn{1}{||c||}{535} & 54 & 51 & 54 & 53 & 54 & 53 & 53 & 53 & 53 & 50 & 50 & 50 & 50 & 50 & 50 & 51 & 53 & 54 & 1471 \\
        \textbf{\# Match-By-Likes}      &  \multicolumn{1}{||c||}{361} & 39 & 44 & 37 & 40 & 42 & 40 & 45 & 45 & 43 & 38 & 40 & 41 & 38 & 38 & 36 & 41 & 41 & 40 & 1089 \\
        \textbf{\# Multiple-Match-By-Likes}      & \multicolumn{1}{||c||}{80} & 8 & 4 & 13 & 7 & 6 & 8 & 7 & 8 & 8 & 6 & 4 & 5 & 7 & 7 & 7 & 8 & 10 & 10 & 213 \\ 
        \textbf{\# Closest-Match-By-Likes}      &  \multicolumn{1}{||c||}{94} & 7 & 3 & 4 & 6 & 6 & 5 & 1 & 0 & 2 & 6 & 6 & 4 & 5 & 5 & 7 & 2 & 2 & 4 & 169 \\
        
        \textbf{Final \# Videos Matched}      &  \multicolumn{1}{||c||}{516} & 51 & 50 & 53 & 52 & 54 & 53 & 53 & 53 & 53 & 49 & 48 & 50 & 50 & 50 & 49 & 49 & 53 & 52 & 1438 \\
        \textbf{\# Unique Videos Per Crawl}      &  \multicolumn{1}{||c||}{390} & 49 & 50 & 53 & 52 & 54 & 53 & 53 & 53 & 53 & 49 & 48 & 50 & 50 & 50 & 49 & 49 & 53 & 51 & 485 \\
        \midrule
        
    \end{tabular}}
    \label{tab:methods_experiment_datasets}
\end{table*}

\parheading{Custom Content Labeling.} We aim to analyze whether the visual and audio content included in videos shown to users in \tkm{} are child-directed and whether they contain advertisements or inappropriate content. As shown in Figure~\ref{fig:auditing_methodology}, Step 2, we conduct a custom content labeling step using the following six content labels that we develop based on COPPA~\cite{ftc_coppa_faq_2020, ftc_youtube_content_rules_2019} and YouTube's documentation~\cite{youtube_child_channels}\footnote{Per the 2019 FTC settlement (also discussed in Section~\ref{subsec:backg_enforcement}), YouTube provides mandatory guidelines for channel owners to indicate whether their videos are child-directed, following closely with COPPA.}.

\begin{enumerate}
    \item[L1]{Does this video contain characters, celebrities, or toys that appeal to children, including animated characters or cartoon figures?}
    \item[L2]{Does this video include activities or content that appeal to children, such as play-acting, simple songs or games, or early education?}
    \item[L3]{Does this video contain or depict any young children?}
    \item[L4]{Does this video contain advertising that is directed to children (\eg{} toys, games for kids)?}
    \item[L5]{Does this video contain advertising of any kind for products or services (\eg{} not specifically child-directed)?}
    \item[L6]{Does this video contain any content (audio, visual) that is inappropriate for children (\ie{} sexual, violent, obscene, scary, or mature content or themes)?}
\end{enumerate}

For each question, the possible answers are ``yes'', ``no'', and ``maybe''. 
Two researchers used these questions to label all the video observations in our dataset manually and discussed throughout the process to reach consensus. Any ``maybe'' labels were discussed and either ``yes'' or ``no'' was selected as the label, and thus the final labels are binary (\ie{} ``yes'' or ``no'') for each question.

Additionally, for L6, we use OpenAI's content moderation model~\cite{openai_moderation} to check for harmful content in the textual data (\eg{} video descriptions, video hashtags, and profile descriptions), which is collected in the metadata extraction step, discussed next. OpenAI's content moderation model analyzes the input text for content across several categories, such as sexual, violent, illicit, and self-harm content, and it outputs scores between 0 and 1 for each category corresponding to whether the input text contains such content, where a higher score indicates higher confidence.

\parheading{TikTok Metadata Extraction.} 
We collect metadata about each video using HTTP requests with each video's URL, which consists of the author's profile name and video ID (\eg{} ``www.tiktok.com/@\textit{profilename}/video/\textit{videoID}''). The responses include metadata regarding the author of the video, such as the description on their profile, summary statistics regarding their profile and videos (\eg{} followers, likes, shares, comments), and their verification status, as well as metadata about the video, including \tiktok{}'s internal content categories (\ie{} referred to as ``diversification labels'' in the response's data structure), whether the video is an advertisement, the video description, and video engagement statistics (\eg{} likes, comments, shares). 
Thus, the result of Step 2 in Figure~\ref{fig:auditing_methodology} is our \dataset{}, complete with our custom content labels, content categories, and both video and video author metadata to enrich our analyses.

\subsection{Experimental Design and Dataset Overview}\label{subsec:method_experiments}

We aim to analyze content curation and the prevalence of child-directed, inappropriate, and advertising content on \tkm{}. We design two kinds of data collection tasks, which result in subsets of our \dataset{}: (1) \neutralcoll{}, in which we collect a large dataset of 500 videos (referred to as ``Neutral Dataset'') that are shown on \tkm{}'s \fyp{} with a single neutral testing account, and (2) \experiments{}, in which we vary the user account ages and usernames (\ie{} with gendered names) across three devices and collect the first 50 videos that are shown to each account on the \fyp{} at the same time. In total, we collected \totobsall{} \tkm{} video observations and manually labeled them, resulting in \totmatchall{} matches (\ie{} we obtained the corresponding video ID). Table~\ref{tab:methods_experiment_datasets} provides detailed dataset statistics.

We conduct the \experiments{} to determine whether different accounts yield different \tkm{} content. We aim to validate the usage of only one testing account in our \neutralcoll{}. Thus, we analyze the content shown to different accounts with varying ages and usernames and find that there is no significant difference across our experiments (see Section~\ref{subsec:results_experiments}), and thus our Neutral Dataset can serve as a representative dataset of what users realistically see on \tkm{}, regardless of their account details. Note that we do not study content personalization in this work. Our experiments are used to validate our data collection approach and provide further insights into \tkm{} content curation, such as video repetition.

\subsubsection{Selecting Account Birthdays and Names}\label{subsubsec:method_exp_accounts}
Our experimental design includes selecting dates of birth and usernames for our user accounts. For the date of birth of all our accounts, we utilized the same day and month, September 9, because it was identified as the most common birthday in the US~\cite{abrams_common_bday_2017}. For the birth year, we used the age grouping guidelines from the United Kingdom (UK)'s Children's Code for age-appropriate design~\cite{gdpr_age_groups_2018}, which are based on children's developmental stages and are as follows: 0--5, 6--9, and 10--12. We select an age in the middle of each grouping (\ie{} 3, 7, and 11) to serve as our three experimental ages, and we then select birth years to match those ages at the time of the experiments. 

For usernames, we utilize the US Social Security Administration (SSA)'s historical data regarding most popular names by year for male and female babies~\cite{ssa_baby_names} to select usernames, matching with the years selected in the previous step. For example, for a female user account in the 6--9 age group, its date of birth selected at the time of this work would be September 9, 2017, and its username would contain the name ``Emma''. To ensure unique usernames during the account creation process\footnote{We develop an automated account creation process, following the same approach as prior work~\cite{kaplan_tiktokaudit_2024}.}, we always append a randomized 10 character string to the selected name.

\subsubsection{\neutralcoll{}}\label{subsubsec:method_neutral_collection}
For our \neutralcoll{}, we create a neutral account where the username does not indicate gender, namely the word ``user'' followed by a randomized 10 character string. We select the middle age group (6--9) for our neutral account, and thus the age is set to seven years old. We only use one account for our \neutralcoll{} to avoid repeated content due to differing accounts and because we observed that account age and username did not make a substantial impact on the content shown, as we observed in our \experiments{} (see Section~\ref{subsec:results_experiments}). Our goal is to create a large dataset of \tkm{} video observations to attempt to characterize the content curation process, and thus we apply our \tkm{} auditing framework to create our neutral account and attempt to collect and match 500 \tkm{} videos. To avoid straining our \tiktokapi{} access and undue load on \tiktok{}'s servers, we implement randomized delays throughout the crawling process, and we only crawl 50 videos at at time. Additionally, since we may come across \tkm{} videos for which there is no match in \tiktok{}'s regular mode, such as due to the author's profile and all their videos being deleted, our actual \tkm{} observations may be more than the number of videos in our final dataset. We complete Step 2 in Figure~\ref{fig:auditing_methodology} to produce a subset of our \dataset{} that is specific to our \neutralcoll{} (\ie{} Neutral Dataset), which contains \totobsneutral{} \tkm{} video observations and \totmatchneutral{} matches. See Table~\ref{tab:methods_experiment_datasets} for our dataset summary.

\subsubsection{\experiments{}}\label{subsubsec:method_age_name_experiments}
For our \experiments{}, we conduct crawls on three separate devices at the same time in two sets of experiments: (1) we vary the user age and keep the account name constant, and (2) we vary the gender indicated in the username and keep the account age constant. Since we have three age groups and three possible usernames (\ie{} female, male, and control (``user'')), our experimental design includes six experimental categories with 18 accounts. 
Our experimental accounts' ages and names are listed in Table~\ref{tab:methods_experiment_datasets}, where ``E'' refers to the experiment set and ``C'' refers to the categories within the experiment. For example, ``E1C1'' in Table~\ref{tab:methods_experiment_datasets} refers to experiment 1, category 1, where we vary the user age (3, 7, and 11) and keep the gendered name constant (``Noah''), whereas ``E2C2'' refers to experiment 2, category 2, in which we vary the users' names (``Emma'', ``Liam'', ``User'') and keep the age constant (7).

Thus, we conduct our experiments by running our crawler script on three separate devices, corresponding to the three experimental accounts, and attempt to collect 50 \tkm{} videos for each account. We run the experiments at the same time across all devices to avoid potential temporal variance in the sequence of videos that may appear. We complete Step 2 in Figure~\ref{fig:auditing_methodology} for each experimental dataset to produce a subset of our \dataset{} specific to our \experiments{}, which contains \totobsexp{} \tkm{} video observations and \totmatchexp{} matches, summed across all experiments. Table~\ref{tab:methods_experiment_datasets} includes more detailed dataset statistics.

\section{Analysis and Findings}\label{sec:findings}

This section discusses our analysis of our \dataset{} and findings, as depicted in Figure~\ref{fig:auditing_methodology}, Step 3, and guided by our RQs in Section~\ref{sec:intro}.

\subsection{Aggregate Analysis (RQ1--2)}\label{subsec:results_agg_analysis}

First, we discuss our analysis of our \dataset{}, which aggregates both our Neutral Dataset (resulting from our \neutralcoll{}) and \experiments{} datasets.

\subsubsection{\tiktok{}'s Content Categorization}\label{subsubsec:results_tiktok_content_categories}

\begin{figure}[t!]
    \centering
    \includegraphics[width=0.4\textwidth]{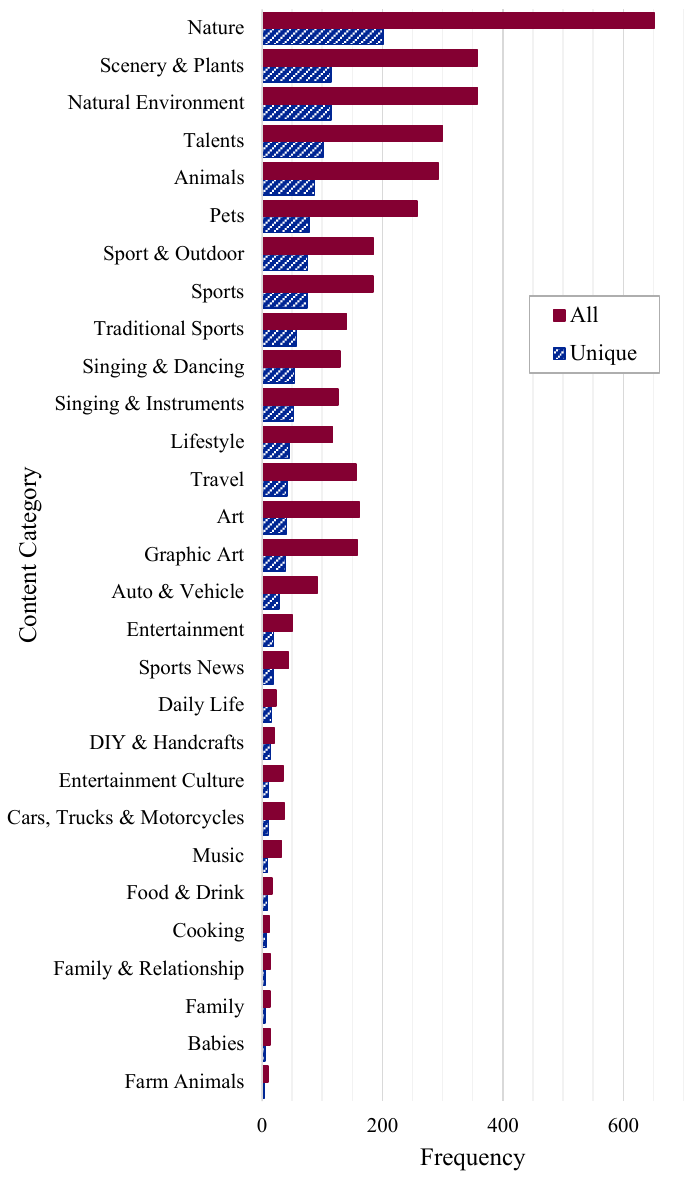}
    \begin{minipage}{\linewidth}
      \caption{
      \textbf{Frequencies of Content Categories Across \dataset{}.} \small{
      This graph visualizes the frequencies of content categories across the unique videos in our \dataset{} and across all observations (\ie{} including repeated observations). We observed 55 content categories, and due to space, we omit 26 categories for which we observed frequencies fewer than 10. See Appendix~\ref{appendix:content_categories} Table~\ref{tab:agg_and_repeat_content_categories_frequencies} for the complete list.
      }}
      \label{fig:content_categories_bar_chart}
    \end{minipage}
\end{figure}

Among our \dataset{}, we observed 55 unique content categories, extracted from \tiktok{}'s metadata per video, as explained in Section~\ref{subsubsec:method_postprocess}. Each video typically has more than one content category, and among the sets of categories associated with each video, we observed 35 unique sets of content categories. Figure~\ref{fig:content_categories_bar_chart} presents a bar graph of the frequencies of the content categories across our \dataset{} corresponding to unique videos (\ie{} referred to in the figure as ``Unique'') and across all the observations without removing any repeated videos (\ie{} referred to in the figure as ``All''). Due to space, we omit 26 low-frequency categories for which we observed fewer than 10 times both among our unique observations and across all observations. We list all the content categories and their frequencies in Appendix~\ref{appendix:content_categories} Table~\ref{tab:agg_and_repeat_content_categories_frequencies}.

The most frequently observed content category, both based on unique \tkm{} videos we observed and based on all the observations in our dataset, is ``Nature'' (\ie{} 201 unique videos and 651 total observations). The next most frequently observed content categories are ``Scenery \& Plants'', ``Natural Environment'',  ``Talents'', ``Animals'', ``Pets'', and ``Sports \& Outdoor''. While these content categories can be related to child-directed content, we observed that there are many content categories that are contextually more related to children's content that appeared very infrequently. Notably, content categories such as ``Anime \& Comics'', ``Toys \& Collectables'', and ``Comics \& Cartoon, Anime'' were observed with very low frequencies (\ie{} less than 10), as shown in Appendix~\ref{appendix:content_categories} Table~\ref{tab:agg_and_repeat_content_categories_frequencies}.
Considering that \tiktok{} has these internal content categories in their metadata, it is peculiar that we observed such low frequencies for categories that are contextually more relevant to children's content and very high frequencies for more generic, seemingly less child-oriented content categories, such as ``Nature'' and ``Scenery \& Plants''.

\subsubsection{Video and Author Prevalence}\label{subsubsec:results_video_author_prev}

In our \dataset{}, we observed \totaggunique{} videos originating from 443 unique video authors, 33 of which were verified accounts~\cite{tiktok_verified_2025}. At the time of this work, these video authors had an average of 
622K followers (minimum 0, maximum 54M), 
19M likes across all their videos (minimum 138, maximum 1.6B), 
and 721 videos on their profiles (minimum 1, maximum 43K). 
Additionally, across the aggregate videos observed, the videos had an average of
48K likes (minimum 0, maximum 2.2M),
273 comments (minimum 0, maximum 6493), and
4410 shares (minimum 0, maximum 207K).
Evidently, the videos come from a very wide variety of video authors based on their prevalence, as shown by the wide range of followers, likes, and number of videos on these profiles. Meanwhile, a majority of the video authors we observed are not verified \tiktok{} users. Additionally, the videos themselves range in engagement, as demonstrated by the likes, comments, and shares statistics in aggregate.

\subsubsection{Custom Content Labeling Results}\label{subsubsec:results_custom_content_labels}

\begin{table}
    \centering
    \caption{\textbf{\dataset{} Custom Content Labeling Statistics. }\small{This table presents statistics regarding our custom content labels among our unique video observations in our \dataset{}, which includes our Neutral Dataset and \experiments{} datasets. ``CD'' stands for ``child-directed'', and ``L\#'' refers to our six custom content labels, as enumerated in Section~\ref{subsubsec:method_postprocess}. L1--4 represent our child-directed content labels, L5 refers to our non-child-directed advertising label, and L6 refers to our inappropriate content label.}}
    \label{tab:agg_content_labels_stats}
    \begin{tabular}{lll}
        \toprule
        \textbf{Category} & \textbf{Count} & \textbf{Proportion} \\
        \midrule
        Not Child-Directed & 401 & 0.83 \\
        Child-Directed (CD), Any 4 Labels & 84 & 0.17 \\
        CD Only L1 (Characters/Celebrities/Toys) & 2 & 0.004 \\
        CD Only L2 (Activities \& Content) & 19 & 0.04 \\
        CD Only L3 (Children Depicted) & 14 & 0.03 \\
        CD Only L4 (CD Advertising) & 0 & 0 \\
        CD At Least L1 & 22 & 0.05 \\
        CD At Least L2 & 67 & 0.14 \\
        CD At Least L3 & 48 & 0.1 \\
        CD At Least L4 & 2 & 0.004 \\
        All 4 CD Labels & 0 & 0 \\
        CD L1, L2, \& L3 & 4 & 0.008 \\
        CD L1, L3, \& L4 & 1 & 0.002 \\
        CD L1, L2, \& L4 & 0 & 0 \\
        CD L2, L3, \& L4 & 1 & 0.002 \\
        CD  L1 \& L2 & 14 & 0.03 \\
        CD  L1 \& L3 & 1 & 0.002 \\
        CD  L1 \& L4 & 0 & 0 \\
        CD  L2 \& L3 & 28 & 0.06 \\
        CD  L2 \& L4 & 0 & 0 \\
        Non-CD Advertising (L5) & 4 & 0.008 \\
        CD \& Non-CD Advertising (L5) & 0 & 0 \\
        Inappropriate Content (L6) & 9 & 0.02 \\
        CD \& Inappropriate Content (L6) & 0 & 0 \\
        \bottomrule
    \end{tabular}
\end{table}

As shown in Figure~\ref{fig:auditing_methodology}, we use our custom content labels to analyze whether the content observed is child-directed, whether it contains inappropriate content, and whether it contains advertisements. Our custom content labeling results are shown in Table~\ref{tab:agg_content_labels_stats}. While our custom content labeling for inappropriate content focuses on the audial and visual components of \tkm{} videos, we also use OpenAI's content moderation model~\cite{openai_moderation} to determine whether there is inappropriate content in the videos' textual descriptions, which are not visible to users in \tkm{} but may provide insight into how \tiktok{} selects content for \tkm{}. To complement our custom advertising content labels, we analyze whether the \tiktok{} metadata discloses advertising content or not for each video. Next, we discuss our findings for each content label:

\parheading{Child-Directed Content.}
As shown in Table~\ref{tab:agg_content_labels_stats}, we observed that 83\% (401/485) of \tkm{} videos were \textit{not} child-directed and 17\% (84/485) were child-directed. 
A video is considered child-directed if at least one of the four labels (L1--4) was labeled affirmatively (\ie{} ``yes''). 
Out of the child-directed videos, we observed
22 with at least L1 in the affirmative (\ie{} contains characters, celebrities, or toys that appeal to children), 
67 with L2 (\ie{} activities or content that appeal to children),
48 with L3 (\ie{} video depicts young children),
and two with L4 (\ie{} child-directed advertising).
Table~\ref{tab:agg_content_labels_stats} also reports the statistics of each individual label and their intersections with other labels.

We also analyze the content categories, provided in \tiktok{}'s metadata, for the child-directed videos. Note that one video typically has more than one content category assigned to it. Among the unique child-directed videos, the most frequent content categories were ``Sports \& Outdoor'', ``Talents'', ``Singing \& Dancing'', and ``Graphic Art''. Examples of child-directed sports-related content we observed were videos of children playing sports, such as in school or sports teams, and examples of child-directed music- and art-related content included videos of people singing songs or drawing characters from popular children's movies.

We did not observe any videos that had all four child-directed labels in the affirmative. The most common combination of affirmative child-directed labels was L2 and L3 (\ie{} 28 videos), which refer to videos that contain activities or content that appeal to children and that depict young children, respectively. This makes sense because the videos that contained activities or content that appeal to children typically depicted children participating in those activities. The second most common combination of affirmative child-directed labels was L1 and L2 (\ie{} 14 videos), which refer to videos that contain characters, celebrities, or toys that appeal to children and activities or content that appeal to children. For example, such videos contained songs or activities (\eg{} drawing) related to child-oriented toys or characters from movies or shows.

We observed two videos that contained child-directed advertising (\ie{} L4), and these videos depicted characters, celebrities, or toys that appeal to children (\ie{} L1), activities or content that appeal to children (\ie{} L2), and/or young children (\ie{} L3), and thus we observed one video with L1, L3, and L4 in the affirmative and one video with L2, L3, and L4 in the affirmative.
Thus, it makes sense that we did not observe any videos that had only L4 in the affirmative.

\parheading{Advertising Content.}
We observed four videos in our \dataset{} that contained advertisements that were not specifically directed to children (\ie{} L5), and as discussed above, we observed two videos that contained child-directed advertisements (\ie{} L4). These six videos were observed across our datasets a total of 12 times. None of these six videos were labeled as advertisements in \tiktok{}'s metadata, and there was no indication on the \tkm{} video that these videos contained advertisements. 
The content categories for these videos include ``Sports'', ``Video Games'', and ``DIY \& Handcrafts''.
These videos originate from distinct \tiktok{} profiles that represent businesses, either digital ones or brick-and-mortar businesses, and one of the accounts is verified.
None of these videos contained any inappropriate content according to our own manual analysis of the audial and visual content (\ie{} L6), nor did they contain any inappropriate textual content in the video descriptions or author profile descriptions, based on OpenAI's content moderation model results~\cite{openai_moderation}.

\parheading{Inappropriate Content.}
We observed a total of nine unique videos on \tkm{} that contained inappropriate content (\ie{} L6), and these videos appeared across our dataset a total of 15 times. The content in these videos that resulted in affirmative inappropriate content labels were music playing in the video that had sexually explicit/suggestive or profane lyrics as well as sexually explicit/suggestive text shown in the video. Regarding the textual descriptions on the videos and on the authors' profiles, we did not find any inappropriate content, based on OpenAI's content moderation model results~\cite{openai_moderation}.
The content categories for these videos include ``Scenery \& Plants'', ``Sports'', ``Singing \& Instruments'', ``Celebrity Clips \& Variety Show'', and ``Pets''.
While the visual content in these videos is largely appropriate, except for the one with sexually explicit text shown on the video, the music playing in the background is often the culprit. Thus, we question whether \tkm{}'s content curation process includes in-depth inappropriate content moderation across multiple modalities, including audio, visual, and text.

\begin{tcolorbox}[title=Key Takeaway 1]
{\bf Observation:} We found that 83\% of observed \tkm{} videos were \textit{not} child-directed, and we observed inappropriate (sexually explicit or profane) content and advertisements, which were not transparently labeled.

{\bf Implication:} \tkm{} is for children by name only and the majority of its videos are actually not child-directed, even including inappropriate content.
\end{tcolorbox}

\subsection{Video Repetition Analysis (RQ1)}\label{subsec:results_video_repeat}

Next, we discuss our findings regarding video repetition in our Neutral Dataset and \experiments{} datasets. Video repetition may cause child users to disengage with \tkm{}, potentially incentivizing them to abandon the platform for riskier alternatives.

\newcommand{\len}{1.27cm}
\newcommand{\lenn}{1.3cm}
\begin{table}[t!]
    \centering
    \small
    \caption{\textbf{Frequencies of Video Repetitions. }\small{This table presents the frequencies of individual video and video sequence repetitions, which we define as a sequence of more than one video repeated more than once, for each of our datasets. Our Neutral Dataset is referred to as ``Neutral''. For the \experiments{} datasets, we refer to each one with the same shorthand as in Table~\ref{tab:methods_experiment_datasets}, and the video sequence repetitions appeared in at least two of the three accounts. The proportion for individual videos repeated is averaged across accounts.}}
    \resizebox{0.4875\textwidth}{!}{
    \begin{tabular}{l cccccc}
    
    \makecell{\textbf{Dataset}} & {\parbox{\len}{\centering{\textbf{\# Videos Repeated}}}} & {\parbox{\len}{\centering{\textbf{Proportion Videos Repeated}}}} & {\parbox{\lenn}{\centering{\textbf{\# of Repeated Seq. (RS)}}}} & {\parbox{\len}{\centering{\textbf{Avg. Length RS}}}} & {\parbox{\len}{\centering{\textbf{Min Length RS}}}} & {\parbox{\len}{\centering{\textbf{Max Length RS}}}} \\
    \midrule

    Neutral & 125 & 24.2\% & 15 & 7.9 & 2 & 17 \\
    E1C1 & 25 & 48.7\% & 2 & 2 & 2 & 2 \\
    E1C2 & 31 & 58.5\% & 4 & 2.2 & 2 & 3 \\
    E1C3 & 43 & 81.1\% & 6 & 2.3 & 2 & 3 \\
    E2C1 & 30 & 61.2\% & 6 & 2.3 & 2 & 4 \\
    E2C2 & 38 & 76.5\% & 7 & 2.4 & 2 & 3 \\
    E2C3 & 34 & 66.3\% & 5 & 2 & 2 & 2 \\

    \midrule
    \end{tabular}
    }
    \label{tab:results_experiments_repeats}
\end{table}

\subsubsection{Neutral Dataset Video Repetition}\label{subsubsec:results_neutral_video_repeat}

Our Neutral Dataset contains 516 videos observations, and 24\% of the videos (\ie{} 125/516) were repeated at least twice, as shown in Table~\ref{tab:results_experiments_repeats}. 
The average distance between repeated video observations was 373 (\eg{} a video was shown once and then repeated 373 videos later), and the median, minimum, and maximum distances between two repeated videos were 376, 14, and 384, respectively.
We also observed 15 repeated videos sequences, meaning a sequence of at least two videos repeated at least twice, and all of these sequences were repeated twice in our Neutral Dataset. The average length of a repeated video sequence was 7.9 videos, and the minimum and maximum lengths were 2 and 17, respectively (\eg{} we observed the same exact 17 videos in a row two times).

For an individual user's account on \tkm{}, they may see content repeated over time, either within one session of watching content on \tkm{} or across multiple sessions, depending on how many videos they watch per session. 
On average, the duration of the videos in this dataset was 30 seconds, and the median, minimum, and maximum durations were 20, 6, and 340 seconds (\ie{} 5.6 minutes), respectively. 
Thus, on average, a user may see a repeated video after three hours of using \tkm{}.
Considering that we observed some large outliers for video duration (\eg{} maximum duration of 340 seconds, or 5.6 minutes), if we use the median video duration, a user may see a repeated video every two hours.

\subsubsection{\experiments{} Video Repetition}\label{subsubsec:results_exp_video_repeat}

Within each Account Age and Username Experiment crawl (\ie{} each of the 18), we observed infrequent individual video repetition---we observed three instances in which a video was repeated twice within one of these crawls. This makes sense in context with our Neutral Dataset findings, where we observed an individual repeated video on average every 373 videos, and our experiment-related crawls collected around 50 videos each. 

However, we observed frequent repetition across accounts within each experimental category. For example, for the E1C3 experiment, as shown in Table~\ref{tab:results_experiments_repeats}, we observed 43 individual videos repeated across all three accounts. Recall that the number of videos observed and matched (\ie{} we identified the video ID) for the E1C3 crawls was 53 for each account (see Table~\ref{tab:methods_experiment_datasets}), and thus 81.1\% of the videos observed in each account were repeated in the other accounts. In Table~\ref{tab:results_experiments_repeats}, we present the average proportion of videos repeated across the three accounts per experiment category, since each crawl resulted in slightly different values for total videos observed (see Section~\ref{subsubsec:method_data_collection} for details). All six experiments resulted in high proportions of repeated videos, where the minimum average proportion was 48.7\%, nearly half (\ie {} for E1C1), and the maximum average proportion was 81.1\% (\ie{} E1C3). All but E1C1 had proportions of repeated videos above 50\%. Thus, across our experiments, nearly half or more of the videos shown across the three accounts within each experiment were the same videos.

We also analyzed repetition of video sequences, which are sequences of more than one video repeated more than once across accounts within an experiment. Table~\ref{tab:results_experiments_repeats} presents the statistics of repeated video sequences across at least two of the three accounts, as we did not find any sequences that were repeated across all three accounts per experiment. E1C1 had the smallest count of 2, whereas E2C2 had the highest count of 7.
Thus, even when users vary their age and username, there is still frequent repetition on \tkm{}, both for individual videos and sequences of videos.

\begin{tcolorbox}[title=Key Takeaway 2]
{\bf Observation:} \tkm{} contains frequent individual video and video sequence repetition---24\% (\ie{} 125/516) of the videos in our Neutral Dataset were repeated at least twice, and a sequence of 17 videos was repeated twice.

{\bf Implication:} Repetitive content will likely cause lack of interest and engagement with \tkm{}, and, when coupled with the lack of child-directed content, may lead to children opting for riskier platforms.
\end{tcolorbox}

\subsection{Content Differences Across Accounts (RQ1)}\label{subsec:results_experiments}
Our \experiments{} datasets include a total of 936 observations across 18 individual account-specific crawls across the six experiments. 
These experiments provide validation for the usage of one testing account for our \neutralcoll{}.

\begin{table}[t!]
    \centering
    \caption{\textbf{Frequencies of Custom Content Labels in \experiments{}. }\small{
    This table presents the frequencies of our custom content labels across videos shown to each account in our \experiments{}. ``CD'' refers to L1--4 for child-directed content, ``Inappr.'' refers to L6 for inappropriate content, and ``NCD Ads'' refers to L5 for non-child-directed advertisements. The frequencies lack substantial variation across the categories, as discussed in Section~\ref{subsec:results_experiments}.
    }}
    \label{tab:fishers_3x3_content_labels}
     \resizebox{0.495\textwidth}{!}{
    \begin{tabular}{p{0.14\linewidth} p{0.005\linewidth} p{0.005\linewidth} p{0.005\linewidth} p{0.005\linewidth} p{0.005\linewidth} p{0.005\linewidth} p{0.005\linewidth} p{0.005\linewidth} p{0.005\linewidth} p{0.005\linewidth} p{0.005\linewidth} p{0.005\linewidth} p{0.005\linewidth} p{0.005\linewidth} p{0.005\linewidth} p{0.005\linewidth} p{0.005\linewidth} p{0.005\linewidth}}
        
         &  \multicolumn{3}{c}{{\textbf{E1C1}}} &  \multicolumn{3}{c}{{\textbf{E1C2}}} &  \multicolumn{3}{c}{{\textbf{E1C3}}} &  \multicolumn{3}{c}{{\textbf{E2C1}}} & \multicolumn{3}{c}{{\textbf{E2C2}}} &  \multicolumn{3}{c}{{\textbf{E2C3}}} \\
        
        \textbf{Account} & A1 & A2 & A3 & A1 & A2 & A3 & A1 & A2 & A3 & A1 & A2 & A3 & A1 & A2 & A3 & A1 & A2 & A3 \\
        
        \midrule
        
        \multicolumn{1}{l|}{\textbf{CD}}      & 9 & 8 & \multicolumn{1}{c|}{7} & 15 & 12 & \multicolumn{1}{c|}{9} & 12 & 13 & \multicolumn{1}{c|}{9} & 6 & 5 & \multicolumn{1}{c|}{5} & 7 & 6 & \multicolumn{1}{c|}{8} & 7 & 6 & 6 \\
        \multicolumn{1}{l|}{\textbf{Inappr.}} & 1 & 1 & \multicolumn{1}{c|}{1} & 1 & 1 & \multicolumn{1}{c|}{1} & 0 & 0 & \multicolumn{1}{c|}{0} & 0 & 0 & \multicolumn{1}{c|}{0} & 0 & 0 & \multicolumn{1}{c|}{1} & 0 & 0 & 0 \\
        \multicolumn{1}{l|}{\textbf{NCD Ads}} & 0 & 0 & \multicolumn{1}{c|}{0} & 1 & 1 & \multicolumn{1}{c|}{0} & 1 & 1 & \multicolumn{1}{c|}{1} & 0 & 0 & \multicolumn{1}{c|}{0} & 0 & 0 & \multicolumn{1}{c|}{0} & 0 & 0 & 0 \\
        
        \midrule
    \end{tabular}
    }
\end{table}

We aim to determine whether \tkm{} account age and username impact the \fyp{}'s videos using content categories (\ie{} provided in \tiktok{}'s metadata) and our custom content labels. For example, for E1C1, we vary the user account's age (\ie{} 3, 7, and 11) and keep the name used in the account's username the same (\ie{} ``Noah'') across all three accounts (see Table~\ref{tab:methods_experiment_datasets}). We analyze whether there are any differences between the content categories and custom content labels of the videos observed across these accounts.

Regarding content categories, we observed 44 individual content categories across our \experiments{} datasets, however more than one content category is typically assigned to the same video (\ie{} see Section~\ref{subsec:results_agg_analysis}), and thus there are 26 unique content category sets. For example, one video may contain the content categories ``Nature'', ``Scenery \& Plants'', and ``Natural Environment''. 
For our statistical test, we choose Fisher's Exact test due to the small frequencies observed for many of the content category sets (\eg{} $<$ 5). We also apply the Freeman-Halton extension (referred to as Fisher-Freeman-Halton) because we are testing three groups (\ie{} accounts) instead of two. 
Thus, we apply the Fisher-Freeman-Halton test on the frequencies of the content category sets across accounts per experiment category. We find \textit{no statistically significant differences} in any of the six experiments at the 0.05 significance threshold. Due to space, we include results for two experiments in Appendix~\ref{appendix:exp_stat_tests} Table~\ref{tab:fishers_3xN_content_categories_e1c1_e2c1}, which shows the frequencies of content category sets across three accounts for each experiment and the p-values from the Fisher-Freeman-Halton test.

We also analyze the differences in frequencies of our custom content labels, as shown in Table~\ref{tab:fishers_3x3_content_labels}, to determine whether there are any significant differences between the frequencies of child-directed, inappropriate, and non-child-directed advertising content shown to different accounts. For this test, we group our four child-directed content labels into one binary label (\eg{} if any of the four labels is affirmative, then the content is deemed child-directed). We apply the Fisher-Freeman-Halton test to each experiment category, and we find \textit{no statistically significant differences} in any of the experiments at the 0.05 significance threshold.

Within the context of our dataset, these results suggest that there is no statistically significant difference in the frequencies of content categories nor custom content labels by account age and username. However, our statistical tests are limited by our dataset size for the experiments (\eg{} around 50 videos per account crawled), and we have relatively low frequencies across the categories and labels, thus impacting the statistical results. Nevertheless, the goal is to determine whether there was any impact to the \fyp{} content from the very beginning of an account with different ages and usernames, and as discussed above, the frequencies were quite similar. We do not aim to study whether the content personalizes over a longer period of time. Such an experiment would require a longitudinal study, which is a future direction (see Section~\ref{subsec:discuss_limitations_future_work}).

\begin{tcolorbox}[title=Key Takeaway 3]
{\bf Observation:} We observed no significant differences between the frequencies of content categories and custom content labels among videos across accounts with varied ages and usernames. 

{\bf Implication:} These findings validate the usage of one testing account for our \neutralcoll{}. They also indicate a lack of age-specific content curation, as is done on similar platforms (\eg{} YouTube Kids~\cite{youtubekids}).
\end{tcolorbox}

\subsection{Insufficient \kidsmode{} App Features (RQ3)}\label{subsec:results_app_features}

As discussed in Section~\ref{sec:methodology} and depicted in Figure~\ref{fig:tiktok_ui_screenshots}, \tkm{} is extremely limited in both the information and features available to children and their parents. Compared to \tiktok{}'s regular mode, \tkm{} restricts features that may pose risks to children, such as in-app direct messaging, comments, the ability to have a public profile (\ie{} all \tkm{} accounts are private and cannot be made public), and posting \tiktok{} videos (\ie{} \tkm{} users can only create and save private videos). However, we find the following critical limitations.

First, there is no parental consent process for children to create a \tkm{} account, as highlighted as a privacy law compliance issue in \cite{figueira_diffaudit_2024}. Thus, parents do not have a way to verify that they are the child's parent at account creation nor associate their email account with their child's \tkm{} account, which are typically required to complete data rights requests, as stipulated in COPPA~\cite{coppa_rule_2013}. The only instance in which \tkm{} prompts the user to involve their parent is when the built-in \tkm{} screen-time limit of one hour has passed, after which a parent is supposed to complete a simple single-digit multiplication problem to be able to set a screen-time passcode, which must be entered to provide more screen time every 30 minutes. Screenshots of the corresponding pages for screen time management are included in Appendix~\ref{appendix:tkm_features} Figure~\ref{fig:tkm_ui_settings_screentime_screenshots}. Not only can this be easily bypassed by a savvy child, this is insufficient for parental controls, as have been studied in prior work regarding parents' privacy and safety control mechanisms (see Section~\ref{sec:related_work}).

Second, several important features are underdeveloped. Particularly, \tkm{} has a button in the ``Settings'' page labeled ``Security'' (see Appendix~\ref{appendix:tkm_features} Figure~\ref{fig:tkm_ui_settings_screentime_screenshots}) that does nothing when pressed, thus there are no ``Security'' options. \tkm{} is missing accessibility features (\eg{} screen reader, text size adjustment, color contrast), which significantly hinders the usability of the app and excludes users with disabilities. There is also a ``Discover'' page which lists many categories of content and corresponding videos, but there is no search functionality to find specific content.

\begin{tcolorbox}[title=Key Takeaway 4]
{\bf Observation:} \tkm{} lacks critical safety, privacy, and usability features, including parental controls, security features, and accessibility options.

{\bf Implication:} \tkm{} is underdeveloped in contrast to \tiktok{}'s regular mode, which contains many of these features, and thus poses safety and privacy risks.
\end{tcolorbox}

\subsection{Comparison to \tiktok{}'s Regular Mode (RQ3)}\label{subsec:results_reg_tiktok_kids_content}

Our findings regarding the lack of child-directed content on \tkm{} lead us to question whether children would be engaged by \tkm{} content enough to stay on the platform. If not, they may venture to \tiktok{}'s regular mode, which is restricted to users 13 years and older. However, children have been known to lie about their age to gain access to such restricted online services~\cite{hargittai2011parents, minkus2015children}, compromising their privacy and safety. Regardless, if children move to \tiktok{}'s regular mode, they will easily encounter child-directed content, either due to \tiktok{}'s recommendation algorithm determining their interests and inevitably showing them child-directed videos, as shown in prior work~\cite{hilbert_bigtech_2024}, or because they can simply search for content that they are interested in---a feature that is not available in \tkm{}. 

Hilbert \etal{}~\cite{hilbert_bigtech_2024} demonstrated through experiments on \tiktok{}'s regular mode that if a user behaves like a child through their interactions, their recommendations will skew towards child-directed content. Additionally, a child does not need to rely on \fyp{} recommendations since it is easy to find child-directed content on \tiktok{} directly. To demonstrate this, we conduct a case study: we manually searched for the most popular and recent television shows and movies for children under 13 on Rotten Tomatoes~\cite{rottentomatoes_kids_movies, rottentomatoes_kids_shows}, a popular media review-aggregation website, as well as popular video games for children on Common Sense Media~\cite{csm_kids_games} and PCMag~\cite{pcmag_kids_games}, both of which conduct media reviews for kids. We search for these media titles on \tiktok{}'s regular mode to see if we can easily find related content.

From our three sources, we extracted 15 television shows and movies and 10 video games for children (\eg{} ``Peppa Pig'', ``Cocomelon'', ``Avatar: The Last Airbender'', the ``Minecraft'' game, and the game ``Crossy Road'')~\cite{rottentomatoes_kids_movies, rottentomatoes_kids_shows, csm_kids_games, pcmag_kids_games}.
We observed that \tiktok{}'s regular mode contained an abundance of videos regarding all 25 media titles. Nearly all (22/25) of the media titles each had at least 10 dedicated profiles posting about that content, and 24 of the 25 had verified accounts associated with them (\eg{} ``@peppapig'', ``@cocomelon''). The availability and volume of these content well surpasses the mere 84 child-directed videos (17\%) we found on \tkm{}, which calls into question \tkm{}'s content curation and lack of child-directed content.

\begin{tcolorbox}[title=Key Takeaway 5]
{\bf Observation:} \tiktok{}'s regular mode contains an abundance of child-directed content that is both easy to find and posted by reputable sources.

{\bf Implication:} \tiktok{} could easily curate content for \tkm{}, and thus it is surprising that only 17\% of the \tkm{} content we observed was child-directed.
\end{tcolorbox}

\section{Discussion}\label{sec:discussion}

In this section, we discuss implications of our findings for children's privacy and safety (\ref{subsec:discuss_implications}), recommendations for children's service providers and regulators (\ref{subsec:discuss_recommendations}), limitations and future directions (\ref{subsec:discuss_limitations_future_work}), and ethical considerations (\ref{subsec:discuss_ethics}).

\subsection{Implications for Children's Safety and Privacy}\label{subsec:discuss_implications}

Our findings have implications for children's safety and privacy. \tiktok{}'s regular mode has been scrutinized for its privacy and safety risks for children, including exposure to harmful content~\cite{ nyag_tiktok_lawsuit_2024, herzlich_nyp_tiktok_meditation_2025, allyn_npr_addiction_known_2024, ftc_tiktok_lawsuit_2024}. \tiktok{} has attempted to solve the problem by creating \tkm{}. However, even though \tkm{} is a restricted version of \tiktok{} conceptualized for children, it lacks critical features for protecting children, and its content is not child-directed. Thus, children may prefer using \tiktok{}'s regular mode instead, where they can be exposed to further risks.

Our work provides insights for parents, service providers, policymakers, and researchers interested in improving children’s digital well-being. To facilitate future research and industry improvement, we plan to publicly release our software artifacts, including the labeled datasets that can aid in improving automated content moderation or conducting user studies, such as evaluating parents' and children's perceptions of the content provided to children. This resource can also support in-depth content analyses and market research by service providers seeking to better understand and address the needs of young users. Our auditing methodology enables longitudinal studies to investigate how content and features on \tkm{} evolve over time and to identify what factors are important for making the content safer and more child-directed.

\subsection{Recommendations}\label{subsec:discuss_recommendations}

\subsubsection{Children's Service Providers}\label{subsubsec:discuss_recommend_providers}

We recommend \tiktok{} and any other child-directed service providers to be more transparent about their content curation process and to implement detailed child-directed content labeling on their platforms. Similarly to the FTC's mandates in their 2019 settlement with YouTube~\cite{ftc_youtube_lawsuit_2019} requiring channel owners to label their child-directed content, \tiktok{} should follow a similar process. \tiktok{} should also further vet such content to ensure only age-appropriate and specifically child-directed content is shown in \tkm{}. Considering our case study findings in Section~\ref{subsec:results_reg_tiktok_kids_content}, there exists a plethora of child-directed content on \tiktok{}'s regular mode, and thus developing an internal child-directed labeling process would enable \tiktok{} to easily identify and curate \tkm{} content.

As discussed in Section~\ref{subsec:results_app_features}, \tkm{} lacks critical safety, privacy, and usability features, such as comprehensive parental controls and accessibility settings. We implore \tiktok{} to develop such features for \tkm{} to provide improved safety and privacy features for children and parents and ensure \tkm{} does not exclude users with disabilities. While \tkm{}'s only current ``parental control'' is a trivial screen-time limit, improved parental controls should include a parental consent process at account creation, options for parents to customize the content categories shown to their child, and a mechanism within the \tkm{} app to perform COPPA data rights requests~\cite{coppa_rule_2013}.

\subsubsection{Regulators}\label{subsubsec:discuss_recommend_regulations}

As discussed in Section~\ref{subsec:backg_enforcement}, the FTC regularly conducts investigations and undergoes enforcement actions against online service providers regarding alleged violations of COPPA. 
This paper did not analyze \tkm{} privacy compliance issues in the context of COPPA---prior work has done this regarding data collection and sharing~\cite{figueira_diffaudit_2024}. Rather, we applied COPPA's definition of child-directed content to analyze the prevalence of such content on \tkm{}, revealing that a large majority of \tkm{} content is \textit{not} child-directed. Because of \tkm{}'s lack of child-directed content and frequent content repetition, which may reduce engagement, these issues may incentivize children to abandon \tkm{} and join \tiktok{}'s regular mode. This risk has privacy compliance implications \wrt{} COPPA, since children would have to lie about their age to gain access and be exposed to non-COPPA compliant data collection and sharing in \tiktok{}'s regular mode. We recommend that regulators investigate this problem and consider both enforcement actions and regulatory changes to stop this questionable behavior.

\subsection{Limitations and Future Work}\label{subsec:discuss_limitations_future_work}

\subsubsection{Dynamic Nature of Social Media}\label{subsubsec:limits_dynamics}
Our auditing methodology is designed to study commercial, black-box platforms such as \tiktok{}, where the underlying algorithms are not disclosed. Our findings are based on observing content available on the platform during our study. However, social media is highly dynamic, and content can be deleted or updated at any time. Our findings remain valid even if content is later deleted, since our goal is to identify patterns inherent to \tkm{}'s operation, rather than to conduct exhaustive content collection. Future work can further explore how platforms such as \tkm{} update content over time and for different geographical jurisdictions.

\subsubsection{Scalability}\label{subsubsec:limits_scalability}
We implemented rate limiting and delays in order to respect \tiktok{}'s bandwidth and avoid bot detection while using the \tiktokapi{}, and thus our data collection is time-intensive. For our \neutralcoll{}, we collected \totobsneutral{} video observations in multiple sessions over several days, amounting to 21.7 hours of data collection. Additionally, for our \experiments{}, we collected \totobsexp{} video observations across six experiments, and each experiment involved three devices collecting videos at the same time. The total time for our experiment data collection is based on the maximum crawl duration per experiment, amounting to 17.4 hours of data collection. On average, it takes 2.7 hours to collect 50 \tkm{} videos using our methodology.
Also, two researchers manually conducted our dataset validation process and custom content labeling on our entire dataset (\ie{} \totobsall{} total observations manually labeled), amounting to approximately 22.7 hours over several days for each researcher.

To improve scalability, future studies can design reliable, automated child-directed content classifiers. However, \tiktok{}'s rate limits still create an upper limit to scalability.
Nevertheless, our dataset is useful for future studies, such as exploring parents' perceptions of child-directed content and expectations of features for child-directed services.

\subsubsection{User Interactions}\label{subsubsec:limits_watch_time}
Due to the nature of our data collection solution, we do not control the amount of time spent watching each video. For each video we observe in a \tkm{} crawl, we must obtain its unique video ID, which we do by crawling the video author's profile in real time to find the corresponding video on their profile based on the number of likes. Thus, the crawler may stay on a video for longer than on other videos if the corresponding profile happens to have a larger number of videos to crawl. However, the goal of this work is not to analyze the impact of user interactions (\eg{} watch time, ``liking'') on content personalization. Rather, we aim to replicate a real user's experience with \tkm{} and investigate the content that appears, regardless of interaction, to characterize the content curated for \tkm{}.

\subsubsection{Generalizability}\label{subsec:discuss_generalize}
In this paper, we implemented our auditing methodology specifically for investigating \tkm{} content. However, our methodology is generalizable to other short-form video content platforms.
Considering \tiktok{}'s success, other service providers have begun incorporating similar short-form video content to their mobile platforms to compete, such as Instagram Reels~\cite{leskin_ig_reels_2020}, YouTube Shorts~\cite{spangler_youtube_shorts_2021}, and even Netflix~\cite{forristal_netflix_2025}. 
If the trend continues and other service providers create child-specific versions of their platforms, similar to \tkm{}, then our methodology can be applied to audit such services as well.

In particular, key components of our methodology can be generalized to other platforms, including programmatic app interactions, data collection and experimental procedures, and custom content labeling (\ie{} child-directed, inappropriate, and advertising content). If other child-directed versions of platforms are similarly limited in features relative to their regular service (\eg{} \tkm{} vs. \tiktok{}'s regular mode), our data collection and validation processes can also be generalized to enable independent auditing of these platforms, specifically removing reliance on app-specific features (\ie{} identifying the original videos by searching for videos with the same number of likes for a given profile, as we did for \tkm{} due to the lack of a ``Share'' button---see Section~\ref{subsubsec:method_data_collection} for details). Future work on other platforms can develop web crawlers to aid in this process, or they can utilize open-source APIs, as we did with the \tiktokapi{}~\cite{teather_tiktok-api_2025}.

\subsection{Ethical Considerations}\label{subsec:discuss_ethics}
To the best of our knowledge, this study does not raise ethical issues. First, neither human subjects nor real users' accounts were involved in our study. Our experiments involved our test \tkm{} accounts interacting with \tkm{}'s user interface programmatically. Second, although our experiments might potentially affect \tiktok{}'s bandwidth, we used a reasonably small number of test \tkm{} accounts and implemented rate limiting and delays.

\section{Conclusion}\label{sec:conclusion}
In this paper, we comprehensively investigate, for the first time, the content offered on \tiktok{}'s \kidsmode{} (\tkm{}) to characterize the content curation and analyze the prevalence of child-directed and age-appropriate content. To that end, we propose a novel auditing methodology and implementation for auditing \tkm{}. We apply our methodology to \tkm{} and find that the majority (83\%) of the platform's content is not child-directed, and there are even videos containing inappropriate content. \tkm{} also lacks important features, including comprehensive parental controls and accessibility options. We discuss the implications of our findings for children's safety and privacy, and we provide recommendations to both children's service providers and regulators to improve protections for children online.

\section*{Acknowledgments}
This work has been supported both by NSF awards FG22924 and FG22490 and a gift from the UC Noyce Initiative. We would also like to thank Bethany Shou-Fu Chang for her help with part of our data labeling process.

\bibliographystyle{IEEEtran}
\bibliography{references}

\appendices

\section*{Appendices}

\section{Content Categories}\label{appendix:content_categories}

\newcommand{\hlen}{2.5cm}
\begin{table}[t!]
    \centering
    \caption{\textbf{Frequencies of Content Categories Across our \dataset{}. }\small{This table presents the frequencies of all content categories in our \dataset{}, where ``Unique'' refers to the unique videos in our dataset and ``Total'' refers to all observations, as discussed in Section~\ref{subsec:results_agg_analysis}.}}
    \label{tab:agg_and_repeat_content_categories_frequencies}
    \resizebox{0.48\textwidth}{!}{
    \begin{tabular}{p{4.5cm} cc}
        \midrule
        \textbf{Category} & {\parbox{\hlen}{\centering{\textbf{Unique Frequency}}}} & {\parbox{\hlen}{\centering{\textbf{Total Frequency}}}} \\
        \midrule
        Nature & 201 & 651 \\
        Scenery \& Plants & 114 & 358 \\
        Natural Environment & 114 & 358 \\
        Talents & 101 & 300 \\
        Animals & 87 & 293 \\
        Pets & 78 & 257 \\
        Sport \& Outdoor & 76 & 185 \\
        Sports & 75 & 184 \\
        Traditional Sports & 57 & 140 \\
        Singing \& Dancing & 54 & 129 \\
        Singing \& Instruments & 52 & 126 \\
        Lifestyle & 45 & 117 \\
        Travel & 42 & 156 \\
        Art & 40 & 161 \\
        Graphic Art & 39 & 158 \\
        Auto \& Vehicle & 28 & 92 \\
        Entertainment & 19 & 51 \\
        Sports News & 18 & 44 \\
        Daily Life & 15 & 23 \\
        DIY \& Handcrafts & 14 & 20 \\
        Cars, Trucks \& Motorcycles & 11 & 37 \\
        Entertainment Culture & 11 & 35 \\
        Music & 9 & 32 \\
        Food \& Drink & 9 & 16 \\
        Video Games & 6 & 7 \\
        Games & 6 & 7 \\
        Cooking & 6 & 12 \\
        Recreation Facilities & 5 & 8 \\
        Babies & 5 & 13 \\
        Family & 5 & 13 \\
        Family \& Relationship & 5 & 13 \\
        Motivation & 4 & 8 \\
        Outfit & 4 & 4 \\
        Farm Animals & 3 & 10 \\
        Work \& Jobs & 3 & 3 \\
        Campus Life & 2 & 6 \\
        Society & 2 & 4 \\
        Food Display & 2 & 3 \\
        Beauty \& Style & 2 & 2 \\
        Finger Dance \& Basic Dance & 2 & 3 \\
        Home \& Garden & 2 & 2 \\
        Diary \& VLOG & 2 & 3 \\
        Cosplay & 2 & 2 \\
        Culture \& Education \& Technology & 2 & 4 \\
        Entertainment News & 1 & 2 \\
        Fitness & 1 & 1 \\
        Drinks & 1 & 1 \\
        Others & 1 & 1 \\
        Street Interviews \& Social Experiments & 1 & 2 \\
        Comics \& Cartoon, Anime & 1 & 8 \\
        Toys \& Collectables & 1 & 1 \\
        Random Shoot & 1 & 1 \\
        Celebrity Clips \& Variety Show & 1 & 1 \\
        Fitness \& Health & 1 & 1 \\
        Anime \& Comics & 1 & 8 \\
        \midrule
    \end{tabular}
    }
\end{table}

In this appendix, we expand on Section~\ref{subsubsec:results_tiktok_content_categories}, which discusses the content categories observed in our \dataset{}.
Table~\ref{appendix:content_categories} lists all 55 unique content categories observed along with their frequencies, both when considering unique videos and all observations.

\section{\experiments{} Findings}\label{appendix:exp_stat_tests}

\begin{table*}[ht!]
    \centering
    \caption{\textbf{Content Category Set Frequencies in \experiments{} E1C1 and E2C1 and Fisher-Freeman-Halton Statistical Test Results. }\small{
    This table presents the frequencies of content category sets observed in \experiments{} E1C1 and E2C1, as well as the p-values resulting from applying the Fisher-Freeman-Halton statistical tests to each experiment. We found no significant differences at the 0.05 threshold. As shown in the table, the frequencies are relatively small and similar across accounts within each experiment. See Section~\ref{subsec:results_experiments} for more details.
    }}
    \label{tab:fishers_3xN_content_categories_e1c1_e2c1}
    \begin{tabular}{l | ccc | ccc}
        
        \multicolumn{1}{c}{} &  \multicolumn{3}{c}{\textbf{E1C1}} &  \multicolumn{3}{c}{\textbf{E2C1}} \\
        \multicolumn{1}{l}{\textbf{Category}} & \textbf{Account 1} & \textbf{Account 2} & \multicolumn{1}{c}{\textbf{Account 3}} & \textbf{Account 1} & \textbf{Account 2} & \textbf{Account 3} \\
        \midrule
        Graphic Art, Art, Talents & 6 & 7 & 8 & 7 & 7 & 7 \\
        Traditional Sports, Sports, Sport \& Outdoor & 4 & 2 & 2 & 7 & 3 & 3 \\
        Scenery \& Plants, Natural Environment, Nature & 10 & 11 & 18 & 15 & 16 & 14 \\
        Sports News, Sports, Sport \& Outdoor & 2 & 0 & 3 & 0 & 0 & 0 \\
        Pets, Animals, Nature & 10 & 8 & 6 & 6 & 8 & 11 \\
        Animals, Nature & 2 & 2 & 1 & 0 & 1 & 1 \\
        Food Display, Food \& Drink, Lifestyle & 1 & 0 & 0 & 0 & 0 & 0 \\
        Singing \& Instruments, Singing \& Dancing, Talents & 6 & 6 & 4 & 4 & 6 & 4 \\
        Travel, Travel, Lifestyle & 2 & 2 & 4 & 4 & 4 & 3 \\
        Finger Dance \& Basic Dance, Singing \& Dancing, Talents & 1 & 1 & 0 & 0 & 0 & 0 \\
        Music, Entertainment Culture, Entertainment & 2 & 2 & 1 & 1 & 1 & 1 \\
        Street Interviews \& Social Experiments, Society, Society & 1 & 1 & 0 & 0 & 0 & 0 \\
        Cars, Trucks \& Motorcycles, Auto \& Vehicle, Auto \& Vehicle & 2 & 2 & 1 & 0 & 0 & 0 \\
        Video Games, Games, Entertainment & 0 & 1 & 0 & 0 & 0 & 0 \\
        Auto \& Vehicle, Auto \& Vehicle & 0 & 1 & 0 & 0 & 0 & 0 \\
        Cooking, Food \& Drink, Lifestyle & 0 & 1 & 0 & 0 & 0 & 0 \\
        Babies, Family, Family \& Relationship & 0 & 0 & 1 & 0 & 0 & 0 \\
        Recreation Facilities, Daily Life, Lifestyle & 0 & 0 & 1 & 0 & 0 & 0 \\
        Random Shoot, Others & 0 & 0 & 1 & 0 & 0 & 0 \\
        DIY \& Handcrafts, DIY \& Handcrafts, Talents & 0 & 0 & 0 & 0 & 0 & 0 \\
        Campus Life, Daily Life, Lifestyle & 0 & 0 & 0 & 1 & 0 & 0 \\
        Comics \& Cartoon, Anime, Anime \& Comics, Entertainment & 0 & 0 & 0 & 1 & 1 & 1 \\
        Farm Animals, Animals, Nature & 0 & 0 & 0 & 1 & 0 & 1 \\
        Motivation, Motivation, Culture \& Education \& Technology & 0 & 0 & 0 & 0 & 0 & 1 \\
        Diary \& VLOG, Daily Life, Lifestyle & 0 & 0 & 0 & 0 & 0 & 0 \\
        Art, Talents & 0 & 0 & 0 & 0 & 0 & 0 \\
        \midrule
        p-value &  & 0.9 & & & 0.9 & \\
        \midrule
    \end{tabular}
\end{table*}

In this appendix, we expand on Section~\ref{subsec:results_experiments}, which discusses our findings regarding the differences between the content shown to accounts in our \experiments{}.
Table~\ref{tab:fishers_3xN_content_categories_e1c1_e2c1} presents the frequencies of content category sets observed within two experiments (\eg{} E1C1 and E2C2) and the p-values obtained from applying the Fisher-Freeman-Halton test. We found no significant differences at the 0.05 threshold.

\section{\tiktok{}'s \kidsmode{} Features}\label{appendix:tkm_features}

\begin{figure*}[t!]
    \centering
    \includegraphics[width=0.9\textwidth]{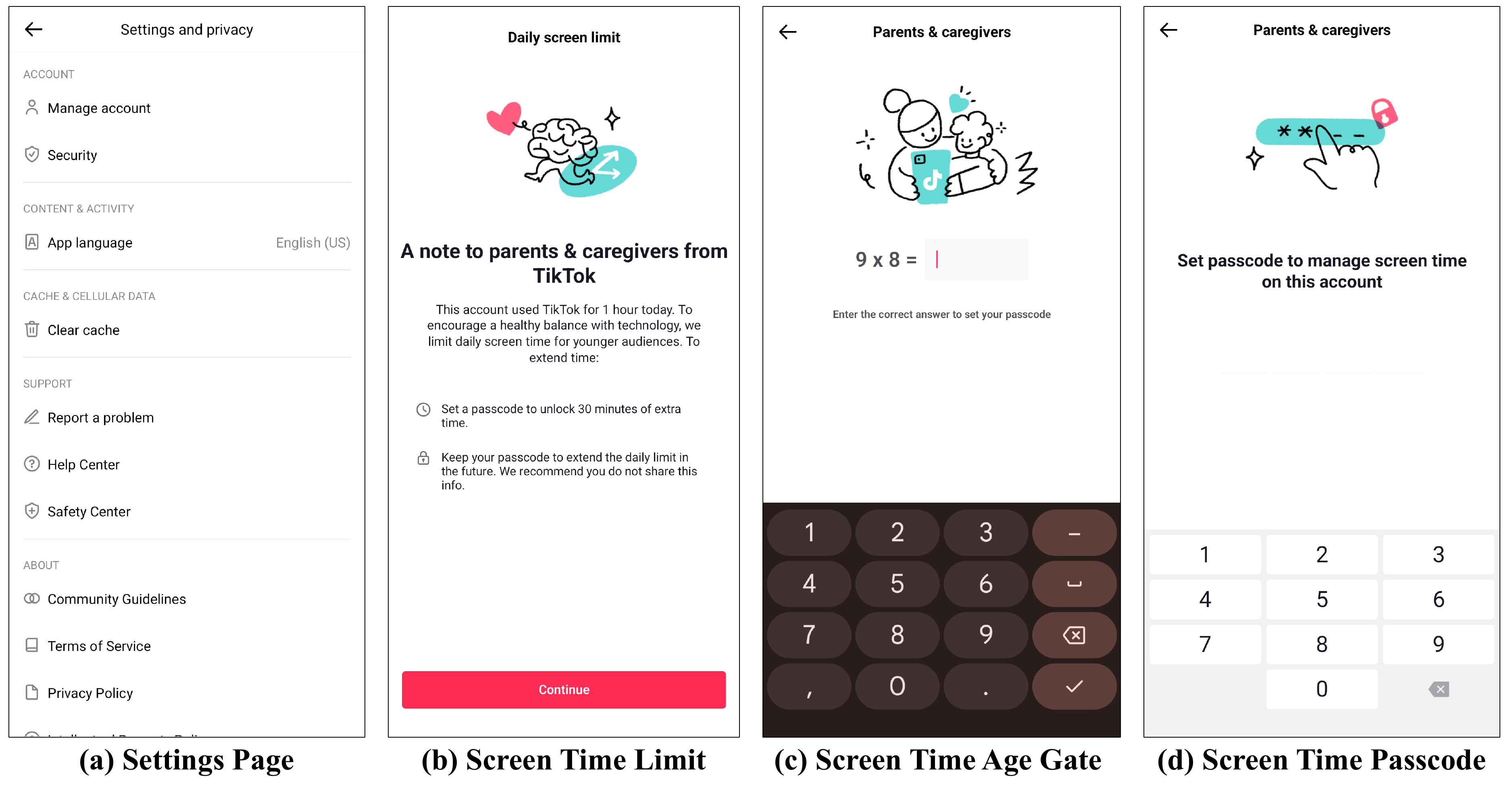}
    \begin{minipage}{\linewidth}
      \caption{
      \textbf{\tiktok{}'s \kidsmode{} Settings and Screen Time Limit Pop-Up Page.} \small{
      This figure presents four screenshots from \tkm{} to expand on the discussion in Section~\ref{subsec:results_app_features}. Figure (a) shows the Settings page, (b) shows the Screen Time Limit page that appears after the first hour of \tkm{} usage has passed, prompting a parent or caregiver to intervene, (c) shows the Screen Time Age Gate page for the parent or caregiver to complete after page (b), and (d) follows (c), prompting the parent or caregiver to create a passcode for screen time management.
      }}
      \label{fig:tkm_ui_settings_screentime_screenshots}
    \end{minipage}
\end{figure*}

In this appendix, we expand on Section~\ref{subsec:results_app_features}, which discusses \tkm{}'s features. In Figure~\ref{fig:tkm_ui_settings_screentime_screenshots}, we provide screenshots of \tkm{} pages to provide context: (a) the Settings page, in which there is a ``Security'' button that does nothing; (b) the Screen Time Limit page, which appears after one hour and then every 30 minutes thereafter; (c) Screen Time Age Gate page, which appears after page (b) and prompts the user's parent or caregiver to complete a simple single-digit multiplication problem; (d) the Screen Time Passcode page appears after (c), prompting the parent or caregiver to set a four-digit passcode to manage screen time.

\end{document}